\documentclass[12pt]{iopart}
\usepackage{graphicx}
\usepackage{graphics}
\usepackage{epsfig}
\usepackage{CJK}
\usepackage{color}
\begin{document}

\title{Transport properties of Li$_{x}$(NH$_{3}$)$_{y}$Fe$_{2}$(Te$_{z}$Se$_{1-z}$)$_{2}$ single crystals in the mixed state}

\author{Shanshan Sun$^{1}$, Shaohua Wang$^{1}$, Chenghe Li$^{1}$, and Hechang Lei$^{1,*}$}

\address{$^{1}$Department of Physics and Beijing Key Laboratory of Opto-electronic Functional Materials $\&$ Micro-nano Devices, Renmin University of China, Beijing 100872, China}
\ead{hlei@ruc.edu.cn}
\begin{abstract}
We study the electric transport properties of Li$_{x}$(NH$_{3}$)$_{y}$Fe$_{2}$(Te$_{z}$Se$_{1-z}$)$_{2}$ single crystals with $z=$ 0 and 0.6 in the mixed state. Thermally-activated flux-flow, vortex glass and flux-flow Hall effect (FFHE) behaviors are observed. Experimental results show that there are rich vortex phases existing in these systems and the vortex liquid states occupy broad regions of phase diagrams. Further analysis suggests that thermal fluctuation plays an important role in the vortex phase diagrams of Li$_{x}$(NH$_{3}$)$_{y}$Fe$_{2}$(Te$_{z}$Se$_{1-z}$)$_{2}$. Moreover, for Li$_{x}$(NH$_{3}$)$_{y}$Fe$_{2}$Se$_{2}$, there is no sigh reversal of FFHE in the mixed state and a scaling behavior $|\rho_{xy}(\mu_{0}H)|=A\rho_{xx}(\mu_{0}H)^{\beta}$ with $\beta\sim$ 2.0 is observed.

\end{abstract}

%Uncomment for PACS numbers title message
%\pacs{00.00, 20.00, 42.10}
% Keywords required only for MST, PB, PMB, PM, JOA, JOB?
%\vspace{2pc}
%\noindent{\it Keywords}: Article preparation, IOP journals
% Uncomment for Submitted to journal title message
%\submitto{\JPA}
% Comment out if separate title page not required
\maketitle

\section{Introduction}

Since the discovery of LaOFeAs\cite{Hosono}, iron-based superconductors (IBSCs) have became another family of high-$T_{c}$ superconductors besides cuprate SCs and attracted intense attention. On the one hand, because of high superconducting transition temperature $T_{c}$ and large upper critical fields $\mu_{0}H_{c2}$, IBSCs are important not only for basic science but also for practical application. On the other hand, the values of Ginzburg number $Gi$ ranging from 10$^{-4}$ to 10$^{-2}$ in IBSCs indicate that the effect of thermal fluctuation varies in different members of IBSCs. It leads to rich vortex phase diagrams in these SCs\cite{Gurevich}.

Among IBSCs, the iron-chalcogenide SC FeSe has a relative low and nearly isotropic $\mu_{0}H_{c2}$ with low $T_{c}$ ($\sim$ 8 K)\cite{HsuFC,Vedeneev}, when compared to iron-pnictide SCs\cite{Hunte}. After intercalating alkali metals A in between FeSe layers using high-temperature synthesis method, the $T_{c}$ can be dramatically enhanced up to about 31 K for A$_{x}$Fe$_{2-y}$Se$_{2}$\cite{GuoJG1}. However, the obvious mesoscopic phase separation between the superconducting phase and the inter-grown antiferromagnetic insulating phase in these compounds make it difficult to study on their intrinsic physical properties\cite{ChenF}. On the other hand, superconducting (Li$_{1-x}$Fe$_{x}$OH)FeSe synthesized by a novel hydrothermal method exhibits the features of high $T_{c}$ ($>$ 40 K), rather large $\mu_{0}H_{c2}$ ($>$ 60 T for $H\Vert c$) and $Gi$ ($\sim$ 1.3$\times$10$^{-2}$)\cite{Lu,Dong,WangZS,YiX}. It results in the existence of various vortex phases in (Li$_{1-x}$Fe$_{x}$OH)FeSe single crystals\cite{YiX,Wen1}.

Besides A$_{x}$Fe$_{2-y}$Se$_{2}$ and (Li$_{1-x}$Fe$_{x}$OH)FeSe, AM-NH$_{3}$ cointercalated FeSe (AM = alkali, alkali-earth, and rare-earth metals) forms another class of iron-chalcogenide SCs with high $T_{c}$ ($>$ 40 K)\cite{Ying,Scheidt,Ying2,Lucas,Sedlnaier}. But the absent of single crystals impedes the study on transport properties and vortex dynamics of these SCs. Very recently, we have grown Li$_{x}$(NH$_{3}$)$_{y}$Fe$_{2}$Se$_{2}$ single crystals successfully and it shows rather high $T_{c}$ and $\mu_{0}H_{c2}$ with significant anisotropy\cite{SSSun}, similar to (Li$_{1-x}$Fe$_{x}$)OHFeSe. In this work, we present a comprehensive study on the electric transport properties of Li$_{x}$(NH$_{3}$)$_{y}$Fe$_{2}$(Te$_{z}$Se$_{1-z}$)$_{2}$ single crystals with $z=$ 0 and 0.6 (denoted as LiFeSe-122 and LiFeTeSe-122 for brevity) in the mixed state. The observed thermally-activated flux-flow (TAFF), vortex glass (VG) and flux-flow Hall effect (FFHE) behaviors indicate that there are diverse vortex phases and rather broad vortex liquid region in these systems. Large $Gi$ in LiFeSe-122 suggests the thermal fluctuation is essential to vortex dynamics. Moreover, the scaling behavior of $|\rho_{xy}(\mu_{0}H)|=A\rho_{xx}(\mu_{0}H)^{\beta}$ with $\beta\sim$ 2 implies that the strength of pinning force is relatively weak in LiFeSe-122 when $H\Vert c$ and temperature is close to $T_{c}$. On the other hand, with Te doping, the $Gi$ decreases accompanying with decreased $T_{c}$.

\section{Experiment}

Single crystals of LiFeSe-122 and LiFeTeSe-122 were synthesized by the low-temperature ammonothermal technique. The detailed experimental procedure and characterizations of crystals were described in previous work\cite{SHWang,SSSun}. The elemental analysis was performed using the inductively coupled plasma atomic emission spectroscopy (ICP-AES) and the atomic ratio of Li$_{x}$(NH$_{3}$)$_{y}$Fe$_{2}$(Te$_{z}$Se$_{1-z}$)$_{2}$ single crystals used in this study is Li : Fe : Se = 0.18 : 1 : 0.9 for $z=$ 0 and  Li : Fe : Te : Se = 0.16 : 1 : 0.60 : 0.38 for $z$ = 0.6, respectively. Electrical transport measurements were performed using a four-probe configuration with current flowing in the $ab$ plane of crystals in a Quantum Design PPMS-14. The Hall resistivity was obtained from the difference of the transverse resistivity measured at the positive and negative fields in order to remove the longitudinal resistivity contribution due to voltage probe misalignment, i.e., $\rho_{xy}(\mu_{0}H)=[\rho(+\mu_{0}H)-\rho(-\mu_{0}H)]/2$.

\section{Results and Discussions}

\begin{figure}[tbp]
\centerline{\includegraphics[scale=0.25]{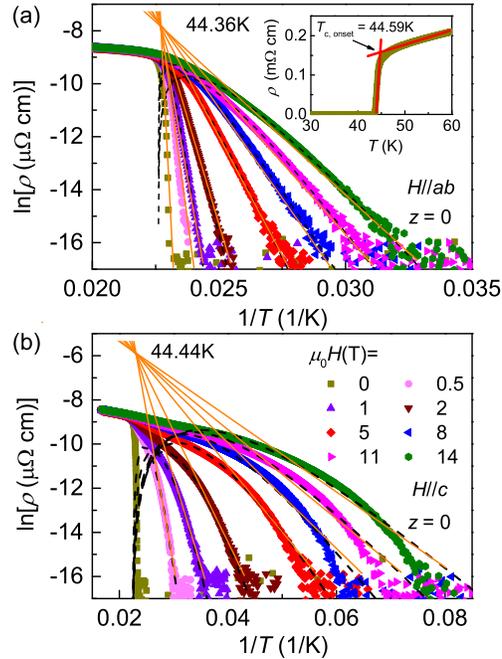}} \vspace*{-0.3cm}
\caption{(a) and (b) The natural logarithm of $\rho(T,\mu_{0}H)$ of LiFeSe-122 as a function of inverse of temperature at various fields for $H\Vert ab$ and $H\Vert c$, respectively. The orange solid and black dashed lines are fitting results from the Arrhenius relation and Eq. (5). Inset : temperature dependence of in-plane resistivity $\rho(T)$ for LiFeSe-122 at zero field.}
\end{figure}

\begin{figure}[tbp]
\centerline{\includegraphics[scale=0.25]{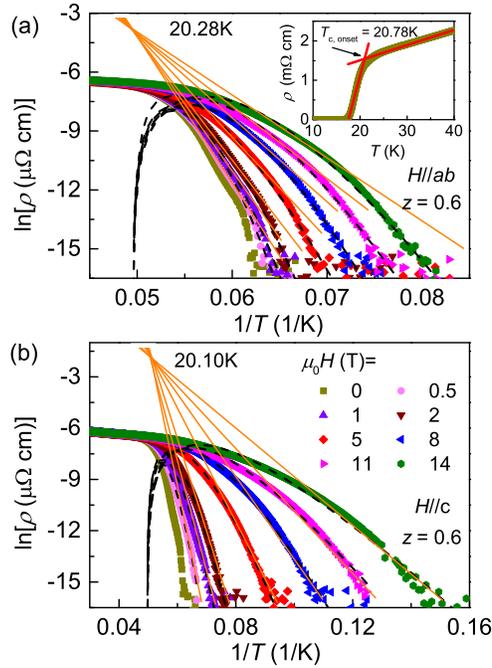}} \vspace*{-0.3cm}
\caption{(a) and (b) ln$\rho(T,\mu_{0}H)$ vs. $1/T$ for LiFeTeSe-122 at various fields with $H\Vert ab$ and $H\Vert c$, respectively. The orange solid and black dashed lines are fitting results from the Arrhenius relation and Eq. (5). Inset : temperature dependence of in-plane resistivity $\rho(T)$ for LiFeTeSe-122 at zero field.}
\end{figure}

Figs. 1 and 2 present the longitudinal resistivity $\rho(T, \mu_{0}H)$ of LiFeSe-122 and LiFeTeSe-122 single crystals near the superconducting transition region for $H\Vert ab$ and $H\Vert c$, respectively. The field-induced broadenings of resistive transitions and resistive tail behaviors are obvious, especially for the field along $c$ axis. Similar behaviors have also been observed in (Li, Fe)OHFeSe, SmFeAsO$_{0.85}$ and cuprates\cite{Dong,Wen1,Lee,Fendrich}. It can be ascribed to the field-induced TAFF. Based on theoretical model describing the TAFF behavior, the resistivity in TAFF region can be expressed as\cite{Palstra1,Blatter,Palstra2},

\begin{equation}
\rho =(2\nu _{0}LB/J){\rm exp}(-J_{c0}BVL/T){\rm sinh}(JBVL/T)
\end{equation}

where $\nu_{0}$ is an attempt frequency for a flux bundle hopping, $L$ is the hopping distance, $B$ is the magnetic induction, $J$ is the applied current density, $J_{c0}$ is the critical current density in the absence of flux creep, $V$ is the bundle volume and $T$ is the temperature. If $J$ is small enough and $JBVL/T\ll 1$, the Eq. (1) can be expressed as

\begin{equation}
\rho =(2\rho _{c}U/T){\rm exp}(-U/T)=\rho _{0f}{\rm exp}(-U/T)
\end{equation}

where $U=J_{c0}BVL$ is the thermally-activated energy (TAE) or flux pinning energy, and $\rho_{c}=\nu _{0}LB/J_{c0}$, which is usually considered to be temperature
independent. According to the condensation model, $U(T, \mu_{0}H)=U_{0}(\mu_{0}H)(1-t)^{q}$, where $t=T/T_{c}$ ($T_{c}$ is the superconducting transition temperature), $q=2-n/2$ ($n$ depends on the dimensionality of the vortex system with the range from 0 to 3) and $U_{0}(\mu_{0}H)$ is the apparent activated energy\cite{Palstra1,Palstra2,Brandt}. Generally assuming $n=2$ and the prefactor $\rho_{0f}$ is a constant as for cuprate SCs\cite{Palstra1}, we can obtain

\begin{equation}
{\rm ln}\rho(T, \mu_{0}H)={\rm ln}\rho_{0}(\mu_{0}H)-U_{0}(\mu_{0}H)/T
\end{equation}
and
\begin{equation}
{\rm ln}\rho_{0}(\mu_{0}H)={\rm ln}\rho_{0f}+U_{0}(\mu_{0}H)/T_{c}
\end{equation}

Thus, there is a linear relationship between ln$\rho(T, \mu_{0}H)$ and $1/T$ in TAFF region (Arrhenius relation) and the slope and $y$-axial intercept correspond to $U_{0}(\mu_{0}H)$ and ln$\rho_{0}(\mu_{0}H)$, respectively. As shown in Fig. 1(a) and (b), the solid lines represent the results of linear fitting in TAFF region using the Arrhenius relation. All the linear fittings cross at $T_{\rm cross}$ approximately, which is about 44.36 K and 44.44 K for $H\parallel ab$ and $H\parallel c $, respectively. Ideally, all the lines at different fields should be crossed into one same point, $T_{\rm cross}$, which should equal to $T_{c}$\cite{Lei2}. Obviously, the values of $T_{\rm cross}$ are consistent with the $T_{c,\rm onset}$ in the $\rho (T)$ curves for both field directions (44.59 K and 44.56 K).

Although the ln$\rho(T, \mu_{0}H)$ vs. $1/T$ can be fitted linearly, there are relative large fitting errors for $H\Vert c$ because the Arrhenius relation can only be satisfied in a narrow region. It suggests that the assumptions of linear temperature dependence of $U(T,\mu_{0}H)$ and temperature-independent $\rho_{0f}$ may not be valid, leading to ln$\rho(T, \mu_{0}H)$ vs. $1/T$ deviating from Arrhenius relation\cite{Zhang YZ2}. Using the relation $U(T,\mu_{0}H)=U_{0}(\mu_{0}H)(1-t)^{q}$, the Eq. (2) can be rewritten as

\begin{equation}
\ln\rho =\ln(2\rho _{c}U_{0})+q\ln (1-t)-\ln T-U_{0}(1-t)^{q}/T
\end{equation}

where $\rho _{c}$ and $U_{0}$ are temperature independent and value of $T_{c}$ is derived from Arrhenius relation. All fits (black dashed lines) using Eq. (5) are also plotted in Fig. 1(a) and (b). It can be seen that this more general method is in better agreement with experimental data than Arrhenius relation, especially for $H\Vert c$.

Same analysis procedure can be applied to LiFeTeSe-122. As shown in Fig. 2 (a) and (b), $T_{\rm cross}$ obtained from the linear fitting is about 20.28 K and 20.10 K for $H\Vert ab$ and $H\Vert c$, respectively. The values of $T_{\rm cross}$ are close to the $T_{c,\rm onset}$ in the $\rho (T)$ curves (20.78 K and 20.79 K). In contrast to LiFeSe-122, the ln$\rho(T, \mu_{0}H)$ of LiFeTeSe-122 for both field directions are not linearly proportional to $1/T$, i.e., the Arrhenius relation can not capture the bending trends of curves. However, the data can be fitted well by using Eq. (5).

\begin{figure}[tbp]
\centerline{\includegraphics[scale=0.24]{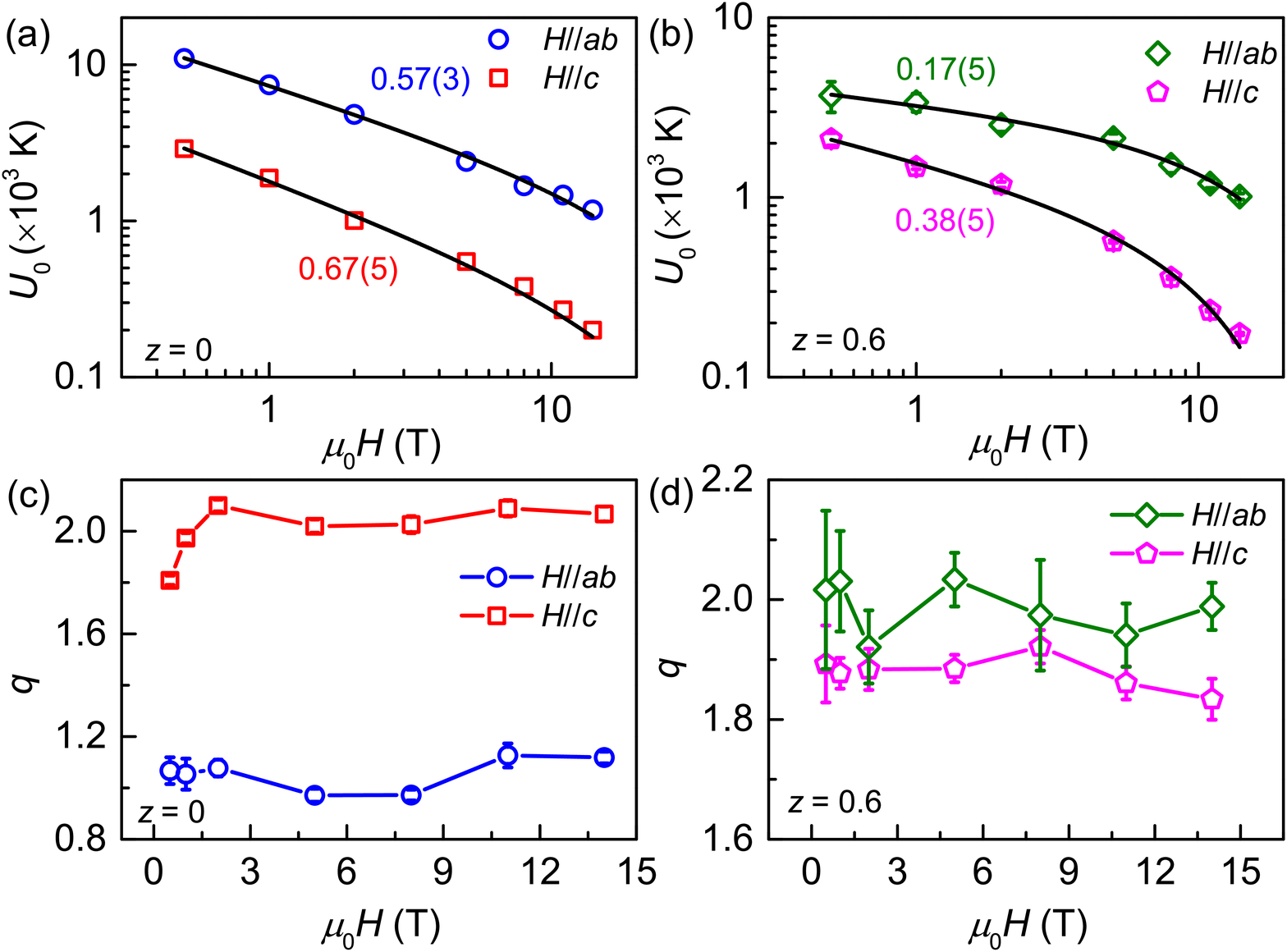}} \vspace*{-0.3cm}
\caption{Field dependence of (a, b) $U_{0}$ and (c, d) $q$ obtained from the fits of the resistivity in the TAFF region using Eq. (5) for LiFeSe-122 and LiFeTeSe-122, respectively. The solid lines in (a) and (b) are fits using Eq. (6) and the values beside the lines are fitted $\gamma$.}
\end{figure}

The fitted $U_{0}$ and $q$ at various fields for both samples are shown in Fig. 3. According to the Anderson-Kim model, $U_{0}(\mu_{0}H)$ is related to the effective pinning energy. $U_{0}(\mu_{0}H)$ for $H\Vert ab$ are much larger than that for $H\Vert c$ in both crystals (Fig. 3(a) and (b)), confirming that the flux pinning is much stronger for $H\Vert ab$. It is consistent with the layered structure of Li$_{x}$(NH$_{3}$)$_{y}$Fe$_{2}$(Te$_{z}$Se$_{1-z}$)$_{2}$. When compared with $U_{0}(\mu_{0}H)\sim$ 250 K at $\mu_{0}H=$ 0.5 T for FeSe\cite{Lei1}, the $U_{0}(\mu_{0}H)$ for $H\Vert c$ at same field enhances to about 10$^{4}$ K in LiFeSe-122, indicating a much stronger vortex pinning energy accompanying with an increased $T_{c}$ after intercalating the Li-NH$_{3}$ layers in between FeSe layers. In contrast, the value of $U_{0}(\mu_{0}H)$ remains in the same order of magnitude (10$^{2}$ $\sim$ 10$^{3}$ K) after intercalating the Li-NH$_{3}$ layers into FeTe$_{1-x}$Se$_{x}$\cite{Shahbazi,Wu}. That is to say, their effective pinning energy are comparable. According to Kramer's law\cite{Thompson}, the field dependence of $U_{0}(\mu_{0}H)$ can be expressed as

\begin{equation}
U_{0}(\mu_{0}H)=a(\mu_{0}H)^{\gamma}(1-\mu_{0}H/\mu_{0}H_{\rm irr})^{\delta}
\end{equation}

where $\mu_{0}H_{\rm irr}$ is an irreversible field and $\gamma$, $\delta$ are scaling parameters. In LiFeSe-122, when assuming $\delta=$ 2, the field is parallel to $ab$ plane, the value of $\gamma$ is 0.57(3) and 0.67(5) for $H\Vert ab$ and $H\Vert c$, respectively (Fig. 3(a)). On the other hand, the value of $\gamma$ is 0.17(5) for $H\Vert ab$ and 0.38(5) for $H\Vert c$ in LiFeTeSe-122 (Fig. 3(b)). The weak field dependence of $U_{0}$ implies that single-vortex pinning should be dominant in LiFeTeSe-122. In contrast, the faster decreases of $U_{0}(\mu_{0}H)$ in LiFeSe-122 imply that the collective flux pinning maybe dominate\cite{Yeshurun}. The better fits using Eq. (5) than Arrhenius relation can be partially ascribed to considering the temperature dependence of prefactor $\rho_{0f}$. When the assumption of $U\gg T$ with $q=$ 1 is satisfied, the Arrhenius relation is valid and it can well describe the TAFF behavior, as in LiFeSe-122 for $H\Vert ab$. But because the obtained $U_{0}$ of LiFeSe-122 for $H\Vert c$ and LiFeTeSe-122 for both field directions are much smaller than that of LiFeSe-122 for $H\Vert ab$, the assumption of temperature-independent $\rho _{0f}$, i. e. $U\gg T$, might be improper and the Arrhenius relation will become invalid.

As shown in Fig. 3(c) and (d), the $q$ for both crystals are almost independent of the field strength for both directions. When $H\Vert c$, the values of $q$ for both samples are about 2, but it changes from 1 in LiFeSe-122 to 2 in LiFeTeSe-122 when $H\Vert ab$. The $q=1$ for $H\Vert ab$ in LiFeSe-122 is consistent with the good linear behavior of ln$\rho(T,\mu_{0}H)$ vs. $1/T$ shown in Fig. 1(a). In contrast, the value of $q=2$ should be another reason leading to the deviation of experimental curves from Arrhenius relation.

According to the VG model\cite{Andersson1,Andersson2}, the linear resistivity disappears as a power law
\begin{equation}
\rho =\rho _{0}|\frac{T}{T_{g}}-1|^{s}
\end{equation}
close to the glass transition temperature, $T_{g}$. In Eq. (7), $\rho_{0}$ is identified as a characteristic resistivity and should in some way be related to the normal state resistivity and $s$ is the glass critical exponent. $T_{g}$ can be extracted by applying the relation $(d\ln\rho/dT)^{-1}\propto(T-T_{g})$ to the resistive tail, and $T^{*}$ is defined as the temperature deviating from the straight line, as shown in Fig. 4.
%Moveover, the normalized resistivity $\rho/\rho_{n}$ versus scaling temperature $t_{s}=(T_{c}-T_{g})/T_{g}(T_{c}-T)-1$ should collapse on a single curve. This scaling behavior can be clearly seen in LiFeSe-122 for $H\Vert c$ (Fig. 4). Similar behaviors have also been observed when $H\Vert c$ and in LiFeTeSe-122 for both field directions (not shown here). It indicates that there are VG regions in these materials, as in (Li, Fe)OHFeSe and BaFe$_{1.9}$Ni$_{0.1}$As$_{2}$\cite{YiX,Su}.

\begin{figure}[tbp]
\centerline{\includegraphics[scale=0.19]{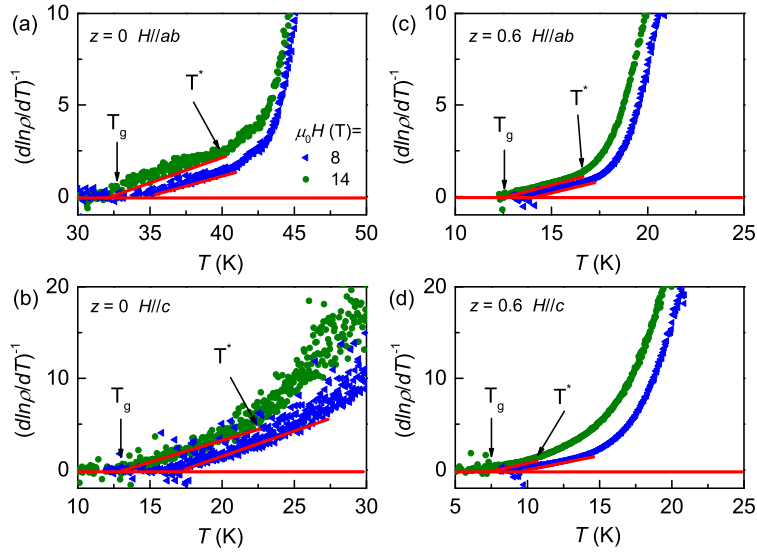}} \vspace*{-0.3cm}
\caption{$(d\ln\rho/dT)^{-1}$ vs. $T$ at $\mu_{0}H=$ 8 and 14 T of (a, b) LiFeSe-122 and (c, d) LiFeTeSe-122 for $H\Vert ab$ and $H\Vert c$, respectively.}
\end{figure}

\begin{figure}[tbp]
\centerline{\includegraphics[scale=0.22]{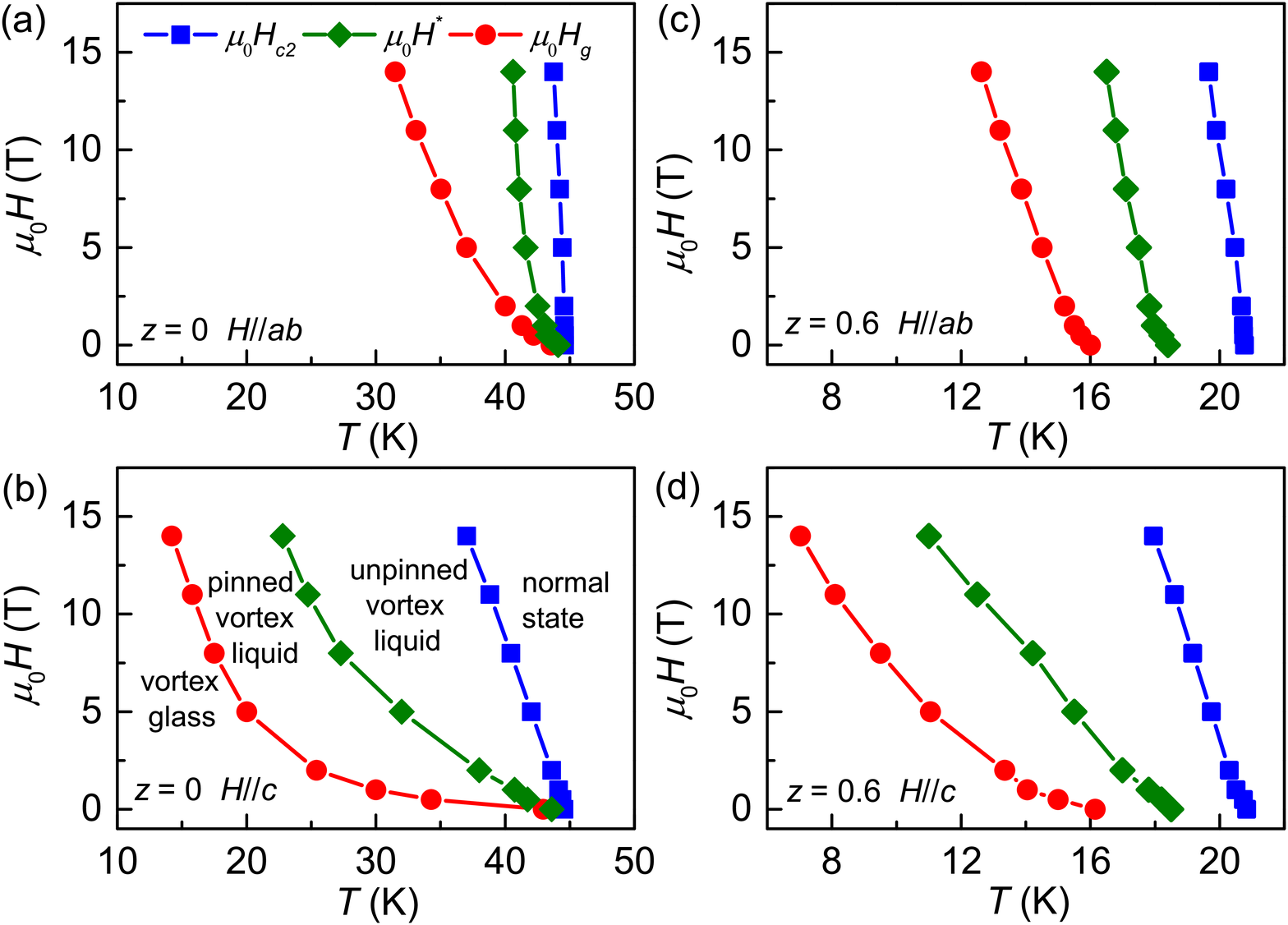}} \vspace*{-0.3cm}
\caption{The vortex phase diagrams ($\mu_{0}H-T$) of (a, b) LiFeSe-122 and (c, d) LiFeTeSe-122 for $H\Vert ab$ and $H\Vert c$, respectively. $\mu_{0}H_{c2}$ is the upper critical field defined from the 90\% $\rho_{n}$; $\mu_{0}H^{*}$ represents the transition from unpinned to strongly pinned vortex region and $\mu_{0}H_{g}$ represents the transition from vortex liquid to glass region.}
\end{figure}

Based on the values of $T_{g}$ and $T^{*}$ extracted from VG model and $\mu_{0}H_{c2}$ defined from the 90\% $\rho_{n}$, we plot the vortex phase diagrams of LiFeSe-122 and LiFeTeSe-122 (Fig. 5). The different regions indicate different flux-pinning mechanism. The region below the $\mu_{0}H_{g}$ line is a vortex glass phase and the region above $\mu_{0}H_{c2}$ is the normal state, while the region between $\mu_{0}H_{g}$ and $\mu_{0}H_{c2}$ is the vortex liquid phase. Besides, $\mu_{0}H^{*}$ represents the divider of unpinned and strongly pinned regions of vortex liquid. Obviously, the vortex liquid region is broad in these crystals, especially in LiFeSe-122 with $H\Vert c$, suggesting a weak vortex-pinning ability even at relative low field in this system when $T$ approaching $T_{c}$.

Because the resistive transition curve $\rho(T,\mu_{0}H)$ not only shifts but also broadens when field increases, it suggests that thermal fluctuation is essential to vortex dynamics in these materials. The strength of thermal fluctuations can be quantified by the $Gi = (\gamma^{2}/2)[(\mu_{0}k_{B}T_{c})/(4\pi B_{c}^{2}(0)\xi_{ab}^{3}(0))]^{2}$\cite{Eley}, where $\gamma$ is the electronic mass anisotropy, $B_{c}(0)=\Phi_{0}/(2\surd2\pi\lambda_{ab}(0)\xi_{ab}(0))$ is the thermodynamic critical field, $\Phi_{0}$ is the flux quantum, $\xi_{ab}(0)$ and $\lambda_{ab}(0)$ is the coherence length and penetration depth at 0 K for $H\Vert ab$. With $\gamma=\xi_{ab}/\xi_{c}$, we obtain $Gi = [(2\pi\mu_{0}K_{B}T_{c}\lambda_{ab}^{2}(0))/(\Phi_{0}\xi_{c}(0))]^{2}/2$, where $\xi_{c}(0)$ is the coherence length at 0 K for $H\Vert c$. Assumed $\lambda_{ab}(0$)= 200 nm same as (Li, Fe)OHFeSe\cite{YiX}, the value of $Gi$ is 1.4$\times10^{-2}$ and 2.8$\times10^{-4}$ for LiFeSe-122 and LiFeTeSe-122 with $\xi_{c}(0)=$ 0.27 and 0.9 nm derived from $\mu_{0}H_{c2}(0)$. The value of $Gi$ for LiFeSe-122 is comparable with (Li, Fe)OHFeSe ($1.3\times10^{-2}$) and YBCO ($10^{-2}$), but much larger than that of LiFeTeSe-122\cite{YiX,Eley}. It is consistent with the more obvious broadening of $\rho(T,\mu_{0}H)$ in LiFeSe-122, especially for $H\Vert c$. The remarkable thermally-activated vortex dynamics in LiFeSe-122 also lead to the wider region of vortex glass/liquid states in vortex phase diagrams when compared to LiFeTeSe-122.

\begin{figure}[tbp]
\centerline{\includegraphics[scale=0.25]{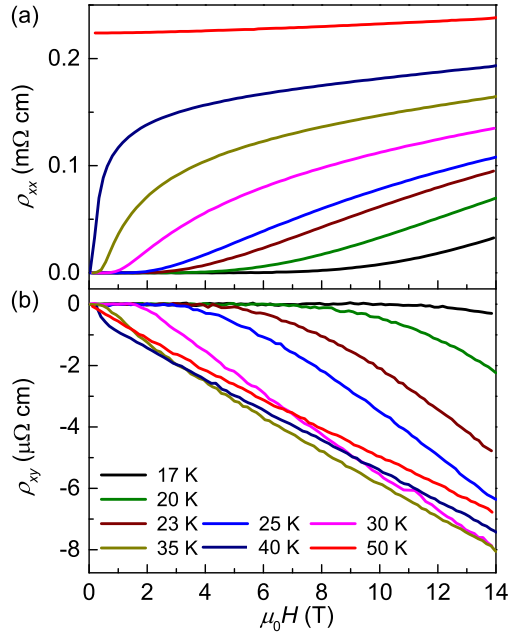}} \vspace*{-0.3cm}
\caption{Field dependence of (a) longitudinal resistivity $\rho_{xx}(\mu_{0}H)$ and (b) Hall resistivity $\rho_{xy}(\mu_{0}H)$ of LiFeSe-122 at various temperatures for $H\Vert c$.}
\end{figure}

The Hall effect in the mixed state is another measurement to study vortex dynamics. Fig. 6(a) shows the field dependence of longitudinal resistivity $\rho_{xx}(\mu_{0}H)$ of LiFeSe-122 in the temperature range of 17 K - 50 K when $H\Vert c$. With increasing field, the superconductivity is suppressed gradually and the transition of $\rho(\mu_{0}H)$ shifts to lower magnetic fields when temperature increases. $\rho_{xx}(\mu_{0}H)$ shows a weak positive magnetoresistance at the normal state ($T=$ 50 K), consistent with previous results\cite{SSSun}. As shown in Fig. 6(b), at the mixed state, the Hall resistivity $\rho_{xy}(\mu_{0}H)$ at low field is zero and becomes negative with increased absolute values at the high fields. The high-field values of $\rho_{xy}(\mu_{0}H)$ gradually reach that in the normal state when temperature is slightly higher than $T_{c}$. The sign of the Hall resistivity is negative, indicating that the electron type carriers dominate in the mixed state as well as in the normal state, consistent with the electron doping in LiFeSe-122. Moreover, there is no sign reversal of $\rho_{xy}(\mu_{0}H)$ in the mixed state, which is often observed in cuprates\cite{Cagigal,Hagen}.

\begin{figure}[tbp]
\centerline{\includegraphics[scale=0.3]{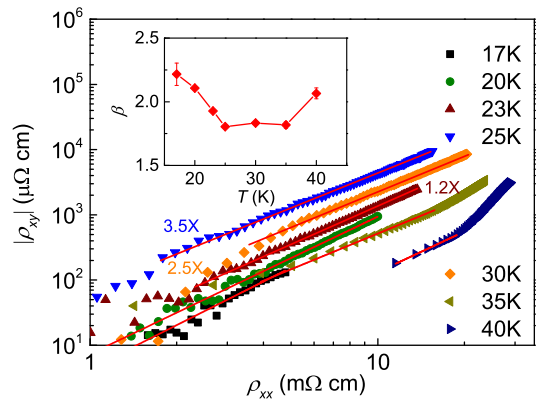}} \vspace*{-0.3cm}
\caption{$|\rho_{xy}|$ vs. $\rho$ at various temperatures for LiFeSe-122. The solid lines are fitting results using the scaling behavior $|\rho_{xy}(\mu_{0}H)|=A\rho_{xx}(\mu_{0}H)^{\beta}$. Inset: temperature dependence of $\beta(T)$.}
\end{figure}

Furthermore, there is a scaling behavior of $|\rho_{xy}(\mu_{0}H)|=A\rho_{xx}(\mu_{0}H)^{\beta}$ in LiFeSe-122 (Fig. 7) and the fitted values of $\beta$ are close to 2 in the whole measuring temperature range. Different values of $\beta$ have been observed in IBSCs and cuprate SCs, such as Fe(Te, S) ($\beta=$ 0.9 - 1.0)\cite{Lei2}, Ba(Fe$_{0.9}$Co$_{0.1}$)As$_{2}$ ($\beta=$ 2.0(2))\cite{WangLM}, and Bi$_{2}$Sr$_{2}$CaCu$_{2}$O$_{y}$ ($\beta$ = 2.0 $\pm$ 0.1)\cite{Samoilov}. A number of theories have been proposed to explain the value of $\beta$. For example, when considering the effect of pinning on the Hall resistivity, Vinokur et al proposed a phenomenological model where $\beta$ = 2.0 in the TAFF region\cite{Vinokur}. And a unified theory for the Hall effect including both the pinning effect and thermal fluctuations was developed by Wang, Dong, and Ting\cite{Wang ZD1,Wang ZD2}. They explained scaling behavior by taking into account the backflow current effect on flux motion due to pinning and pointed out $\beta$ changing from 2 to 1.5 for increasing pinning force. The exponent $\beta\sim$ 2.0 and the absence of sign change of $\rho_{xy}(\mu_{0}H)$ for LiFeSe-122 single crystals imply that the strength of pinning force is relatively weak when $H\Vert c$ and $T\rightarrow T_{c}$, consistent with the low TAE at same field direction shown above.

\section{Conclusion}

In summary, the field-induced resistive broadenings of superconducting transitions and resistive tail behaviors have been observed in Li$_{x}$(NH$_{3}$)$_{y}$Fe$_{2}$(Te$_{z}$Se$_{1-z}$)$_{2}$ single crystals. The detailed analysis shows that these results are associated with the TAFF and VG behaviors. Moreover, the thermal fluctuation plays an important role in the vortex dynamics of Li$_{x}$(NH$_{3}$)$_{y}$Fe$_{2}$(Te$_{z}$Se$_{1-z}$)$_{2}$ single crystals, especially for LiFeSe-122, resulting in broad vortex liquid regions in the vortex phase diagrams. The FFHE with $\beta\sim$ 2 indicates the relatively weak vortex pinning force in LiFeSe-122 when magnetic field is along the $c$-axis and $T$ approaches $T_{c}$.

\ack{
This work was supported by the Ministry of Science and Technology of China (2016YFA0300504), the National Natural Science Foundation of China (No. 11574394), the Fundamental Research Funds for the Central Universities, and the Research Funds of Renmin University of China (RUC) (15XNLF06, 15XNLQ07).}

\end{document}